\documentclass[reprint,aps,pre,amsmath,amssymb,superscriptaddress,showpacs]{revtex4-1}
\usepackage{graphicx,float}

\begin{document}

\title{From subexponential distributions to black swan dominance}

\author{Alexei Vazquez}
\email{alexei@nodeslinks}
\affiliation{Nodes \& Links}

\begin{abstract}
The shape of empirical distributions with heavy tails is a recurrent matter of debate. There are claims of a power laws and the associated scale invariance. There are plenty of challengers as well, the lognormal and stretched exponential among others. Here I point out that, with regard to summation invariance, all what matters is they are subexponential distributions. I provide numerical examples highlighting the key properties of subexponential distributions. The summation invariance and the black swan dominance: the sum is dominated by the maximum. Finally, I illustrate the use of these properties to tackle problems in  random networks, infectious dynamics and project delays.
\end{abstract}

\maketitle

\section{Introduction}

Power law distributions have fascinated generations of researches for decades. Cluster sizes at percolation \cite{stauffer79}. Avalanche sizes in self-organized criticality \cite{bak87}. The degree distribution of scale-free networks \cite{barabasi99}. However, when faced with empirical data, calling a power law distribution is on the eye of the beholder \cite{stouffer05, clauset09, malevergne11, broido19, serafino21}. The power law claims are challenged by other distributions. Most commonly the lognormal and stretched exponential distributions \cite{laherrere98, malevergne05, stouffer05}.

The correct distribution is important if scale invariance is all what matters. That property is unique to power law distributions. There may be other questions where having a heavy tail is sufficient. The distribution being power law or lognormal a matter of degree. One such case is the sum of independent and identically distributed random variables $S_n = \sum_{i=1}^n X_i$. In this context we encounter a different type of invariance, the distribution invariance to summation.

The work on stable distributions have stablished the types of distributions that are fully invariant to summation \cite{feller50}. In that class we find the normal distribution. For several distributions, bounded or with superexponential tails, the distribution of $S_n$ converges to the normal distribution when $n\gg1$. The other members of this class are distributions with power law tails. Examples include the Cauchy and L\'evy distributions, with probability density functions $f(x) = 1/\pi(1+x^2)$ and $f(x)=(1/\sqrt{2\pi}\exp(-1/2x)/x^{3/2}$ respectively. For all distributions with power law tails the distribution of $S_n$ converges to the corresponding member in the stable distribution class when $n\gg1$.

The relaxation of summation invariance to the distribution tail sets the stage for a more general class, {\em the subexponential distributions}, with a tail heavier than any exponential \cite{foss13}. Subexponential distributions are the focus of this work. To get there I will go over the central limit theorem in section \ref{section:clt}. Then in section \ref{section:exponential} I will show how it fails for the sum of random variables with an exponential distribution. In sections \ref{section:SLT} and \ref{section:MLT} I review the mathematics work on subexponential distributions. I emphasize key limit theorems and provide some numerical examples that are missing from the math literature. In section \ref{section:blackswan} I apply these limit theorems to risk analysis. I introduce a relative definition of black swans and point out why they should be the center of our attention. In section \ref{section:applications} I go over some applications of the limit theorems of subexponential distributions. Including aggregate connectivity, the giant component transition in random networks, the long time behavior of infectious dynamics and duration uncertainty in activity networks.

\begin{figure}[t]
\includegraphics[width=3.3in]{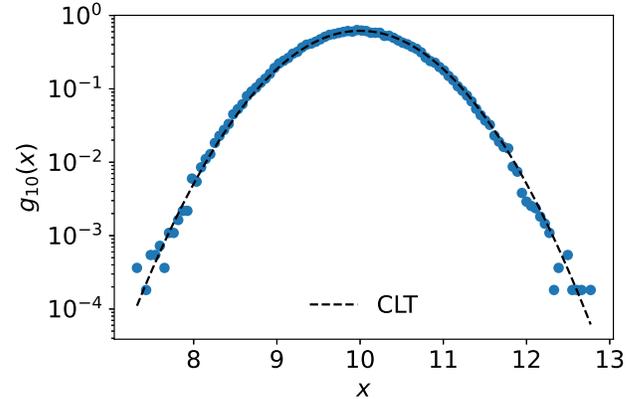}
\caption{Probability density function of the sum of 10 random variables from a triangular distribution with left, mode and right values 0.5, 1 and 1.5 respectively. The dashed line is the normal distribution deduced from the central limit theorem (CLT).}
\label{fig_triangular}
\end{figure}

\section{The central limit theorem}
\label{section:clt}

The sum of random variables
\begin{equation}
S_n = X_1 + \cdots + X_n,
\label{sum}
\end{equation}
is found in many problems. The random variables $X_i$ may be daily changes of a stock price or activity delays in a construction project. The standard text book procedure is to invoke the central limit theorem. If the $X_i$ are independent random variables from the same probability density function $f(x)$ then the probability density function of $S_n$, denoted by $g_n(x)$, is approximated by a normal density
\begin{equation}
g_n(x) = \frac{1}{\sqrt{2\pi n\sigma^2}}\exp\left(-\frac{(y-n\mu)^2}{2n\sigma^2} \right).
\label{normal}
\end{equation}
where $\mu$ and $\sigma^2$ are the expectation and variance of $f(x)$.

For example, the triangular distribution is used to model activity duration uncertainty in construction projects. In turn, the sum of uncertainty along critical activities is used to estimate the project duration uncertainty. It is an intuitive choice. One provides 3 estimates of activity duration (optimistic, expected, pessimistic) and one gets an estimate of the project finish date uncertainty. For this particular case the central limit theorem works perfectly (Fig. \ref{fig_triangular}). This is not an academic example. It is the standard choice in risk analysis for construction projects. A mandatory requirement for modern software in this sector, such as Oracle Primavera, Microsoft Project, Safran and our own platform at Nodes \& Links.

\begin{figure}[t]
\includegraphics[width=3.3in]{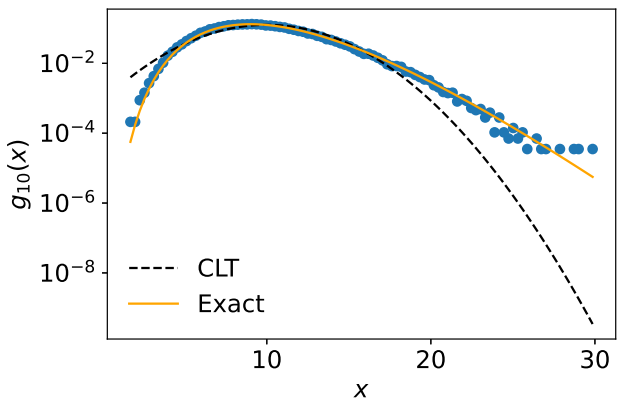}
\caption{Probability density function of the sum of 10 random variables from an exponential distribution with scale parameter $\mu=1$. The dashed line is the normal desnity deduced from the central limit theorem (CLT). The solid line is the exact gamma density in Eq. (\ref{gamma}).}
\label{fig_exponential}
\end{figure}

\section{The exponential distribution}
\label{section:exponential}

The idyllic picture in Fig. \ref{fig_triangular} does not hold in general. The central limit theorem is central for a reason. For the purpose of illustration, consider the exponential distribution. It has the probability density function $f(x) = e^{-x/\mu}/\mu$, where $\mu$ is the scale parameter . In this particular case we calculate $g_n(x)$ exactly, obtaining the gamma density
\begin{equation}
g_n(x) = \frac{1}{\mu (n-1)!} \left(\frac{x}{\mu}\right)^{n-1}  \exp\left(- \frac{x}{\mu} \right).
\label{gamma}
\end{equation}
The gamma density has a maximum (the mode) at $x_n = (n-1)\mu$. If we perform the Taylor expansion of Eq. (\ref{gamma}) around the mode and take the limit $n\gg 1$ we obtain the normal density in Eq. (\ref{normal}). This Taylor expansion is valid in the vicinity of $x\approx x_n$. More precisely for $(x-n\mu)^2\ll n\sigma^2$. The case $n=10$ is shown in Fig. \ref{fig_exponential}. In the vicinity of the peak there is a good agreement between the normal density (dashed line) and the numerical estimate (solid circles). However, there is an evident deviation for small and large values. The central limit theorem does not apply away from the center.

Figure \ref{fig_exponential} corroborates that the numerical estimate falls on the gamma distribution in Eq. (\ref{gamma}). This gamma distribution has an exponential tail, the same as $f(x)$. This is taking us towards a new direction. The invariance of the tail shape.
 
 \section{Subexponential limit theorem}
\label{section:SLT}
 
The work on subexponential distribution shifts the attention from the mode to the tail. We are interested in positive defined variables and therefore the right tail. We define the tail distribution of $f(x)$ and $g_n(x)$ as
\begin{equation}
\bar{F}(x) = {\rm Prob.}\{X>x\} = \int_x^\infty f(\xi)d\xi,
\label{xtail}
\end{equation}
\begin{equation}
\bar{G}_n(x) =  {\rm Prob.}\{S_n>x\} = \int_x^\infty g_n(\xi)d\xi.
\label{ytail}
\end{equation}
The tail of $\bar{G}_2(x)$ has the lower bound
\begin{equation}
\bar{G}_2(x)\geq 2\bar{F}(x)
\label{bound}
\end{equation}
for $x\rightarrow\infty$ \cite{foss13}. The tail of two identically independent random variables is at least as heavy as $2\bar{F}(x)$. For example, for an exponential distribution $f(x)=\exp(-x/\mu)/\mu$, we get $x \geq 2\mu$, which is certainly true for $x\rightarrow\infty$

Subexponentials are distributions with a slow decaying tail that cannot be bounded by any exponential. There is no $\mu>0$ such that $\bar{F}(x) < \exp(-x/\mu)$ for $x\rightarrow\infty$.
For subexponential distributions,  the lower bound in Eq. (\ref{bound}) becomes an approximation \cite{foss13}
\begin{equation}
\bar{G}_2(x) \approx 2\bar{F}(x)
\label{subexp}
\end{equation}
for $x\rightarrow\infty$. The distribution tail of two identically independent random variables with a subexponential distribution is as heavy as $2\bar{F}(x)$. The property (\ref{subexp}) is used as an alternative definition of subexponentiality.

Equation (\ref{subexp}) can be extended to any $\bar{G}_n(x)$ with $n\geq2$. That was proven by Chistyakov in 1964 \cite{chistyakov64}. If $f(x)$ is a subexponential distribution, as defined by Eq. (\ref{subexp}), then
\begin{equation}
\bar{G}_n(x) \approx n\bar{F}(x),
\label{SLT}
\end{equation}
for $x\rightarrow\infty$. Upon summation the tail gets heavier by a factor of $n$, but the shape remains the same as that of $\bar{F}(x)$. I will call this the subexponential limit theorem.

\onecolumngrid

\begin{figure}[t]
\includegraphics[width=7in]{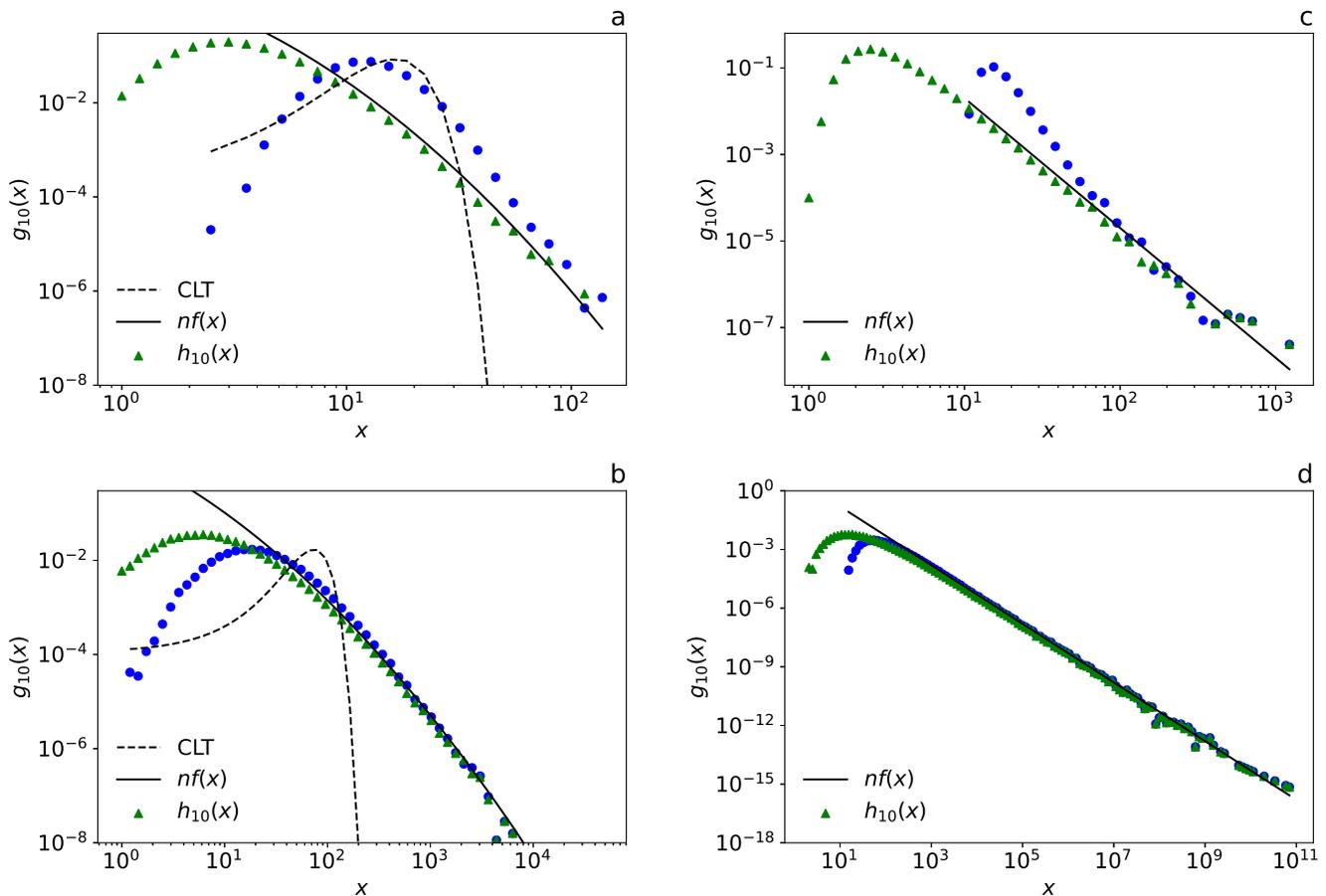}
\caption{Probability density function of the sum of 10 random variables from a subexponential distribution (circles). a) Lognormal, $\mu=0$ and $\sigma=1$. b) Lognormal, $\mu=0$ and $\sigma=2$. c) power law, $\alpha=2$ and $\mu=1$. d) power law, $\alpha=0.5$ and $\mu=1$. The dashed line is the central limit theorem. The solid line is the subexponential limit theorem. The triangles are the numerical estimate of the probability density function of the maximum of the 10 random variables.}
\label{fig_subexponential}
\end{figure}

\twocolumngrid

\subsection{Lognormal distribution}

The lognormal probability density function is
\begin{equation}
f(x) = \frac{1}{x\sqrt{2\pi\sigma^2}} \exp\left( - \frac{(\ln x - \mu)^2}{2\sigma^2}\right),
\label{lognormal}
\end{equation} 
where $\mu$ and $\sigma^2$ are the location and variance of $\ln x$. The tail of the lognormal density is of the order of $\exp( -(\ln x)^2/2\sigma^2)/x$ and decays slower than any exponential. The larger is $\sigma$ the heavier is the tail. The sum of random variables with a lognormal is not approximated by the normal density of the central limit theorem (Fig. \ref{fig_subexponential}a,b, solid circles vs dashed line). In contrast, the subexponential limit $nf(x)$ is a good approximation for the tail of $g_n(x)$ (Fig. \ref{fig_subexponential}a,b, solid circles vs solid line). The agreement expands over a wider range the higher is the degree of subexponentiality. Higher $\sigma$ in this case.

\subsection{Power law distributions}

A Pareto or power law distribution has the probability density function
\begin{equation}
f(x) = \left\{\begin{array}{ll}
0, & {\rm if}\ x<\mu\\
\frac{\alpha}{\mu} \left( \frac{\mu}{x} \right)^{1+\alpha}, & {\rm otherwise}
\end{array}\right.
\label{pareto}
\end{equation}
where $\alpha$ is the tail exponent and $\mu$ is scale parameter. The tail of the power law density is slower than any exponential. The tail is heavier the smaller is $\alpha$.  The power law distribution is scale invariant. If we apply the scale transformation $x\rightarrow ax$ the distribution is transformed as $f(x) \rightarrow f(ax) d(ax) = f(x) dx / a^{\alpha}$. The shape of the distribution remains invariant. Scale invariance is specific to power law densities. Since the power law distribution is subexponential the tail distribution is also invariant to summation. Figures \ref{fig_subexponential}c,d show that indeed, $g_n(x)\approx nf(x)$ for large $x$. Here again the agreement extends to a wider range the higher is the degree of subexponentiality. Lower $\alpha$ in this case.

\section{Maximum limit theorem}
\label{section:MLT}

The subexponential limit theorem has a follow up associated with the maximum. Let
\begin{equation}
M_n = \max\left( X_1, X_2, \ldots, X_n \right),
\label{max}
\end{equation}
be the maximum of independent and identically distributed random variables, with tail distribution $\bar{H}_n(x) = {\rm Prob.}\{M_n>x\}$. For $x\rightarrow\infty$ we have $\bar{F}(x)\rightarrow0$ and therefore (Ref. \cite{foss13}, Definition 3.1) 
\begin{equation}
\bar{H}_n(x) = 1 - (1-\bar{F}(x))^n \approx n \bar{F}(x).
\label{maxlim0}
\end{equation}
Note this limit is valid for any $\bar{F}(x)$, independently of the subexponential property. Now, from Eqs. (\ref{maxlim0}) and (\ref{SLT}) it follows that
\begin{equation}
\bar{G}_n(x) \approx \bar{H}_n(x).
\label{MLT}
\end{equation}
The tail distribution of the sum is approximated by the tail distribution of the maximum. I will call this result the maximum limit theorem.

Figure \ref{fig_subexponential} shows that the tail of $h_n(x)$ is a good approximation for the tail of $g_n(x)$, the probability density function of the sum $S_n=X_1+\cdots X_n$ (circles vs triangles).

\section{Black swan dominance}
\label{section:blackswan}

For the ancient Romans {\em niger cycnus} was quintessential {\em rara in terris} \cite{puhvel84}. Today black swan is a common term for rare events of extreme magnitude. Black swan events are associated with the tail statistics. The heavier the tail of the distribution the higher the chance to observe a large event, a {\em black swan}. There is debate on whether black swans are extreme events in the tail of the distribution or unforeseen events happening at a higher frequency than expected from the reference statitics \cite{demarzo22}.

I propose the maximum limit theorem as a working definition of black swans. If ${\rm Prob.}\{ X_1 + X_2 > x \} \approx{\rm Prob.}\{ \max( X_1 , X_2 ) > x \}$ then the only way $X_1+X_2$ is large is that $\max( X_1 , X_2 )\gg\min( X_1 , X_2 )$. Using Eq. (\ref{MLT}) we can extend that argument. The only way $\sum_i X_i$ is large is that $\max_i X_i  \gg \max_{i\neq k} X_i$, where $k={\rm argmax}_i X_i$ is the largest event index. One event, the black swan, dominates all. That is the definition of {\em black swan dominance}.

The black swan is defined in relative terms. We do not specify a threshold or scale. We demand that $\max_i X_i  \gg \max_{i\neq k} X_i$. Therefore a black swan is context dependent. We can have a construction project where all activities are not delayed, except one delayed by 30 days. Since $30\gg 1$ days we call that a black swan. In another construction project we can have 10 critical activities delayed by about 30 days, resulting in an aggregated delay of 300 days. Since $30\ll 300$ we don't call any of these 30 days delays a black swan. The stress is in the dominance. One event of magnitude much larger than the others.

\section{Applications}
\label{section:applications}

\subsection{Aggregate connectivity}

Suppose we have a network ${\cal N}$ composed of $N$ nodes and $M$ links between pairs of nodes. The degree of a node is the number of links containing that node or, equivalently, the number of neighbors or the number of interacting partners. Let $p_k$ be the degree distribution across nodes. We want to know the aggregate number of links between a community ${\cal C}$ of $n$ nodes and the complement ${\cal N}\setminus {\cal C}$. The $n$ nodes may represent a community in a social network or a pathway in a biological network, for example.

We can approximate the aggregate connectivity by the sum in Eq. (\ref{sum}), where $X_i$ are the nodes degrees. This approximation holds if (i) there is a small overlap between the direct neighbors of the $n$ nodes in ${\cal N}\setminus {\cal C}$ and (ii) the number of links between the $n$ nodes is much smaller than $S_n$. Most real networks, including social and biological networks are characterized by heavy tailed degree distributions \cite{newman_watts_strogatz_02, barabasi_oltvai04}. If $p_k$ has a subexponential tail then the probability density function of $S_n$ has a tail approximated by $n p_k$. Botton line, the aggregation of nodes makes the tail heavier by a factor of $n$. The probability to find extreme hubs with large aggregate connectivity increases by a factor of $n$.

\subsection{Giant component of random networks}

Consider a random network with given degree distribution $p_k$ and excess degree distribution $q_k = kp_k/\langle k\rangle$. $p_k$ characterizes the degree of a vertex selected at random and $q_k$ the degree distribution of a vertex at the end of link selected at random. Molloy \& Read \cite{molloy98} demonstrated a transition from disconnected clusters to a giant component when
\begin{equation}
\theta =\sum_k (k-1)q_k = \frac{ \langle k(k-1)\rangle }{ \langle k\rangle } = 1.
\label{theta}
\end{equation}
This result can be recapitulated using the subexponential limit theorem in Eq. (\ref{SLT}).

The cavity method provides estimates for networks with a tree like structure \cite{callaway00, cohen01, vazquez_weigt03}. In a nutshell, we focus on a node $i$ at the end of a link selected at random. The excess degree is the set of links emanating from node $i$ excluding the link we came from. Node $i$ will have excess degree $k-1$ with probability $q_k$. At this point we want to calculate the probability distribution $Q_k(s)$ that we can reach $s$ other nodes following the excess links of node $i$, given that node $i$ has degree $k$. If the network has a tree like structure then how many nodes are reached from node $i$ can be written as the sum of $k-1$ neighbors plus how many nodes can be reached from each neighbor 
\begin{equation}
S_k = k-1 + X_1 + X_2 \cdots + X_{k-1}.
\label{Sk}
\end{equation}
We note that $S_k$ is a random variable extracted from $Q_k(s)$ and $X_i$ is a random variable extracted from $\langle Q\rangle(s) = \sum_k q_k Q_k(s)$.

If  $\langle Q\rangle(s)$ is a subexponential distribution then $X_1 + X_2 \cdots + X_{k-1}$ is approximately distributed as $(k-1) \langle Q\rangle(s)$ for $s\rightarrow\infty$. In this case
\begin{equation}
Q_k(s) = (k-1) \langle Q\rangle (s-k+1)
\label{QkQ}
\end{equation}
Taking the expectation over $q_k$ in both sides and the limit $s\rightarrow\infty$ we obtain
\begin{equation}
\langle Q\rangle(s) = \theta \langle Q\rangle (s)
\label{QQ}
\end{equation}
This identity holds for $\theta = 1$. Therefore, $\langle Q\rangle(s)$ is subexponential only when $\theta=1$. This is the critical point first obtained by Molloy \& Reed \cite{molloy98} and later extended to percolation on random networks \cite{callaway00,cohen01}. For $\theta<1$, the less connected phase, $Q(s)$ is bounded by an exponential decay. For $\theta>1$, the more connected phase, the deviation from subexponentiality can only be explained by the presence of a giant component. 

This example illustrates the use of the subexponential property in the context of networks with a tree-like structure. We did note invoke scale invariance or a power law distribution of components size. We just demanded the subeponential property, the invariance with respect to summation of the distribution tail.

\subsection{Spreading dynamics}

The spreading of an infectious agent (disease, computer virus, rumor, etc) in static or temporal networks can be modeled as a branching process \cite{vazquez_dimacs04}. The average number of new infectious at time $t$ is approximated by
\begin{equation}
n(t) = R_0 \sum_{n=1}^D R^{n-1} g_n(t),
\label{nt}
\end{equation}
where $R_0$ is the expected reproductive number of the starting node (patient zero), $R$ is the expected reproductive number of any other node in the transmission tree, $D$ is the maximum number of generations (the diameter for static networks) and $g_n(t)$ is the distribution of the sum of $n$ generation times.

The generation time is the time elapsed from acquiring to transmitting the disease. Let $f(t)$ be the probability density function of generation times. If the infection spreads at a constant rate $\mu$, a Poisson process, then $f(t)$ is an exponential distribution and $g_n(t)$ is given by the gamma distribution in Eq. (\ref{gamma}). Substituting Eq. (\ref{gamma}) into (\ref{nt}) we obtain
\begin{equation}
n(t) = R_0 \exp\left( - \frac{t}{\mu} \right) R_0 \sum_{n=1}^D \frac{1}{(n-1)!} \left( \frac{Rt}{\mu}\right)^{n-1} .
\label{ntexp}
\end{equation}
When $t\ll \mu D/R$ the sum approximates the Taylor expansion of the exponential and $n(t)\approx R_0 \exp(-(1-R)t/\mu)$. An exponential decay provided $R<1$. When $t\gg \mu D/R$ the sum is dominated by the $r=D$ term and $n(t)\propto t^{D-1}\exp(-t/\mu)$ \cite{vazquez_polynomial06}. Again an exponential decay for large $t$.

In contrast, if the generation time distribution is subexponential, then according to the subexponential limit theorem, Eq. (\ref{SLT}), $g_n(t) \approx n f(t)$ in the limit $t\rightarrow\infty$. Substituting this result in Eq. (\ref{nt}) we obtain
\begin{equation}
n(t) \approx f(t) R_0 \sum_{n=1}^D n R^{n-1} ,
\label{ntsub}
\end{equation}
The long time limit of infection dynamics with a subexponential distribution of generation times is characterized by the same subexponential tail. In particular, this holds true when $f(t)$ has a power law tail, as previously reported \cite{min11}.

 The time between sexual intercourses follows a power law decay \cite{vazquez_dimacs04}. Therefore, the subexponetial decay is relevant for the infection dynamics of sexually transmitted diseases. The distribution of the inter-event time between submission of two consecutive emails by an email user has a fat tail \cite{barabasi05}. Interestingly, whether the tail is a power law or lognormal was subject to debate \cite{stouffer05}. The marriage of these two distributions into the subexponential class closes the debate. The subexponetial decay is relevant for the infection dynamics of email worms and other infectious agents transmited via email. Finally subexponential distributions of inter-event times are found in many other systems \cite{vazquez_et_al06}.

\subsection{Activity networks dynamics}

Human projects are organized as activity networks, where nodes represent activities and arcs represent logical constraints: predecessor must finish before the successor starts. In the planning phase a project schedule is generated based on estimates of activity durations. A critical path is stablished, containing a sequence of activities from beginning to end that are executed without spare time between them. For paths outside the critical path there is some spare time (float). The total float between two activities, denoted by $T_{ij}$, is the spare time along the path of minimum spare time starting at $j$ and ending at $i$, with a negative sign (consumes delay). If there is no path then $T_{ij}=-\infty$, indicating that no delay will be transmitted from $j$ to $i$.

Once the project starts there are delays in the activity completions. If the delay at an activity exceeds the spare time between the activity an a successor then a delay is transmitted to the successor, initiating a delay cascade. There are two sources of delay. Exogenous delays increasing activity duration independently of what happened to the predecessors, with magnitude $z$. Endogenous delays carried from predecessors, with magnitude $h$. The total delay an activity will experience is $S_2=z+h$. If the distribution of delays is subexponential then, invoking the maximum limit theorem, $S_2\approx\max(z,h)$. In this case the delay $d_i$ in the completion of activity $i$ is approximated by \cite{vazquez_tropical22}
\begin{equation}
d_i \approx \max_{j=1}^N (h_j + T_{ij}),
\label{delay}
\end{equation}
where $N$ is the number of activities and $h_i$ the exogenous delay at activity $i$.

Equation (\ref{delay}) has no reference to delay interactions at intermediate activities in the paths from $j$ to $i$. We simply aggregate the contribution of each exogenous delay independently and then take the max. When the exogenous delay follow a subexponential distribution, their contribution to delays at downstream activities can be calculated independently of each other. Just take the max at the end.

\section{Conclusions}

Researchers dealing with heavy tailed statistics should pay attention to the mathematical literature on subexponential distributions. This class includes, among others, power law, lognormal and stretched exponential distributions. From the practical point the subexponential distributions have two key properties: the distribution tail is invariant to summation and for large values the sum is approximated by the max. Here I have shown some examples were subexponentiality is the only thing that matters. These results are not limited to power law distributions. Debates on whether some empirical data is better described by a power law, lognormal or stretched exponential distribution may be irrelevant for certain problems. 

\section*{Acknowledgements}

This work was partly supported by European Union’s Horizon 2020 and the Cyprus Research \& Innovation Foundation under the SEED program (grant agreement 0719(B)/0124). Nodes \& Links provided support in the form of salary for AV, but did not have any additional role in the study design, data collection and analysis, decision to publish, or preparation of the manuscript.

\bibliographystyle{apsrev4-1}

%\bibliography{risk.bib}

\bibliography{blackswan.bbl}

\end{document}